\documentclass[
reprint,
superscriptaddress,
amsmath,
amssymb,
aps,
longbibliography,
pra,
showpacs,
floatfix
]{revtex4-1}
\usepackage{hyperref}
\usepackage{graphicx}
\usepackage{xcolor} 
\usepackage{booktabs}

\usepackage{tabularx}
\usepackage{multirow}

\usepackage{braket}

\usepackage{comment}
\usepackage{enumitem}
\usepackage{soul}

\definecolor{color1}{rgb}{0,0.25,0.70}
\hypersetup{colorlinks=true,
    linkcolor={color1},
    citecolor={color1},
    urlcolor={color1}
}

\begin{document}

\title{Chiral-phonon-induced anomalous Hall effect in bilayer MoSe$_2$}

\author{Omer Yaniv}
\email{yaniv@mail.tau.ac.il}
\affiliation{School of Physics and Astronomy, Tel Aviv University, Tel Aviv 6997801, Israel}

\author{Swati Chaudhary}
\affiliation{The Institute for Solid State Physics, The University of Tokyo, Kashiwa, Chiba 277-8581, Japan}

\author{Carolina Paiva}
\affiliation{School of Physics and Astronomy, Tel Aviv University, Tel Aviv 6997801, Israel}

\author{Dominik M. Juraschek}
\email{dominik.juraschek@unifr.ch}
\affiliation{School of Physics and Astronomy, Tel Aviv University, Tel Aviv 6997801, Israel}
\affiliation{Department of Applied Physics and Science Education, Eindhoven University of Technology, 5612 AP Eindhoven, Netherlands}
\affiliation{Department of Physics, University of Fribourg, CH-1700 Fribourg, Switzerland}
\affiliation{Paul Scherrer Institut, CH-5232 Villigen PSI, Switzerland}

\date{\today}

\begin{abstract}
We demonstrate that optically driven chiral phonons induce an anomalous Hall response in a nonmagnetic, centrosymmetric material. Using first-principles calculations to parameterize an effective $\mathbf{k}\cdot\mathbf{p}$ Hamiltonian for bilayer MoSe$_2$, we show that circular excitation of an $E_u$ phonon generates a circular dynamical Rashba spin-orbit coupling. The resulting Floquet bands carry same-sign Berry curvature at the two valleys, producing a Hall response whose sign follows the phonon handedness. 
We obtain a range of $\sigma_{xy}\sim10^{-2}-0.21$ $\frac{e^2}{2\pi\hbar}$ for experimentally accessible light pulses. Our results show that chiral phonons provide a route to controlling electronic Berry curvature and transport on ultrafast time scales.
\end{abstract}

\maketitle

\section{Introduction}

The anomalous Hall effect (AHE) is a transverse charge response that occurs in the absence of an external magnetic field~\cite{karplus:1954,haldane:2004,nagaosa:2010}.
Its intrinsic contribution is a fundamental manifestation of the
geometric properties of electronic wave functions: the Berry curvature of Bloch bands generates an anomalous transverse velocity, and its integral over occupied electronic states determines the Hall conductivity~\cite{sundaram:1999,Jungwirth2002AHE,Fang2003Monopoles,Yao2004AHE,xiao:2010}. Under time reversal, the Berry curvature changes sign, so that the net anomalous Hall conductivity vanishes in a time-reversal-symmetric system. A finite anomalous Hall conductivity therefore requires broken time-reversal symmetry, often due to magnetic order present in the system~\cite{nagaosa:2010}.
Time-reversal symmetry breaking can alternatively be caused by irradiating a material with circularly polarized light~\cite{oka:2009,lindner:2011,Kitagawa2011Transport,PerezPiskunow2014Floquet}, dressing electronic states into Floquet bands with finite Berry curvature absent in equilibrium~\cite{Shirley1965Floquet,Sambe1973Quasienergies,Bukov2015HighFrequency,wang:2013,rudner:2020,merboldt:2025}. This approach has enabled light-induced gap opening and time-reversal symmetry breaking in the surface electronic states of a topological insulator~\cite{wang:2013}, as well as the observation of light-dressed electronic states in graphene despite its short electronic coherence time~\cite{merboldt:2025}. In transport experiments, circularly polarized light has induced an ultrafast anomalous Hall response in graphene, with contributions from the two valleys adding constructively~\cite{mciver:2020,sato:2019,lesko:2026}.

Light-driven lattice vibrations offer an alternative route to controlling electronic properties by modifying the crystal structure itself~\cite{Forst2011Nonlinear,Subedi2014Nonlinear,juraschek:2017,Disa2021}. Circularly polarized phonons, also known as chiral or axial phonons~\cite{Zhang2015Chiral,Zhu2018Chiral,Juraschek2025}, carry angular momentum~\cite{Zhang2014AngularMomentum} and are intrinsically connected to magnetic responses~\cite{nova:2017,Juraschek2017Dynamical,juraschek:2019,Geilhufe2021Magnetism,juraschek:2022,luo:2023,Basini2024Multiferroicity,Davies2024UltrafastBarnett,Klebl2025Geometry,Shabala2025Axial,Zalewski2026CrystalField,Shabala2026Unifying,Biggs2026ChiralPhononPolaritons,Huang2026Axial,Bilyk2026Domains,Allington2026Spin,Choi2026Ionic} and orbital magnetization~\cite{Xiao2021,Ren2021TopologicalMagnetization,Shabala2024Inverse,Wang2025PhononMoment,Yao2026Orbital,Sato2026Orbital}. At the same time, coherently driven phonons are predicted to generate Floquet bands with the prospect of inducing topological electronic responses~\cite{Iadecola2013DrivenGraphene,Hubener2018Floquet,Shin2018SpinFloquet,Chaudhary2020PhononFloquet,Yang2024Acoustoelectric,Yao2026PhononFloquet}. Inversion-symmetry-breaking polar phonons have further been proposed to induce dynamical Rashba-Dresselhaus spin-orbit coupling in centrosymmetric materials~\cite{schlipf:2021}. These properties of optical phonons motivate our central question: Can circular excitation of phonons produce an anomalous Hall effect similar to circularly polarized light?

Here, we demonstrate a chiral-phonon-induced anomalous Hall response in centrosymmetric 2H-stacked bilayer MoSe$_2$. We show that circular excitation of an $E_u$-symmetric mode produces a circular dynamical Rashba spin-orbit coupling. Using Floquet theory for a $\mathbf{k}\cdot\mathbf{p}$ Hamiltonian with parameters computed from first principles, we find emerging Berry curvature with the same sign at the $K$ and $K'$ valleys. Their contributions produce a Hall response whose sign reverses with phonon handedness.
Our results identify circular dynamical Rashba coupling as a mechanism for converting the angular momentum of coherent lattice motion into electronic Berry curvature and transverse charge transport.

\begin{figure}[t]
    \centering
    \includegraphics[width=0.85\columnwidth]{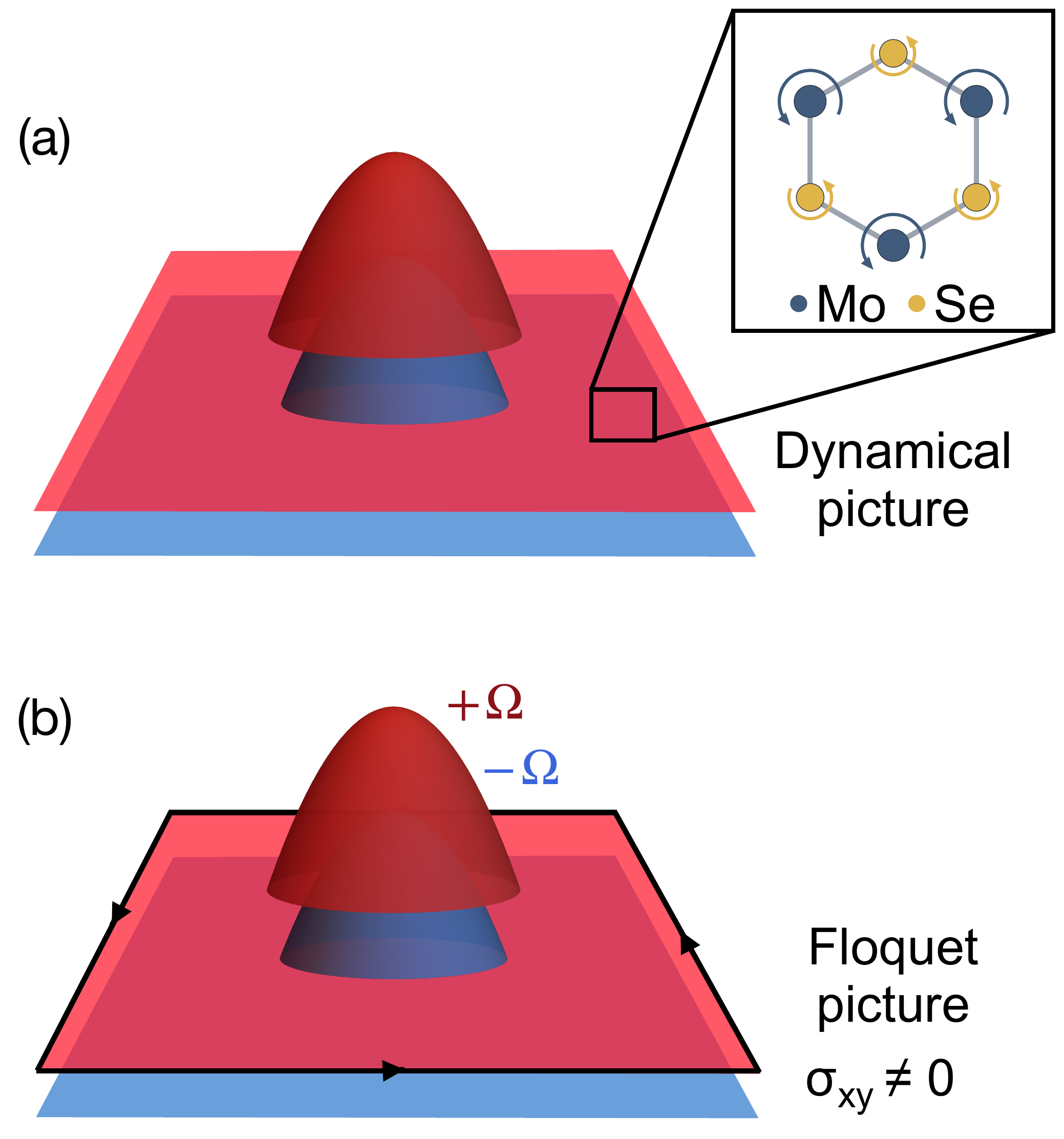}
    \caption{
    \textbf{Illustration of the chiral phonon-induced
    anomalous Hall effect.}
    (a) Dynamical picture: circular excitation of two orthogonal phonon components in a centrosymmetric bilayer produces atomic motion (inset) and a circular dynamical Rashba spin-orbit coupling. (b) Floquet picture: the resulting Floquet bands carry finite Berry curvature and support an anomalous Hall response. Reversing the handedness of the circular motion reverses the sign of the induced Berry curvature.
    }
    \label{fig:mechanism}
\end{figure}

\section{Theoretical formalism}
\label{sec:mechanism}

We start by illustrating the concept of the chiral-phonon-induced anomalous Hall effect, shown schematically in Fig.~\ref{fig:mechanism}. We consider a centrosymmetric, nonmagnetic system with a degenerate electronic doublet. In equilibrium, inversion and time-reversal symmetry enforce a vanishing anomalous Hall conductivity. An inversion-symmetry-breaking phonon can lift the electronic degeneracy through dynamical Rashba spin-orbit coupling~\cite{schlipf:2021}. For a linearly polarized phonon, the displacement oscillates along a fixed direction, and the resulting coupling between electronic spin and momentum does not introduce a definite handedness or break time-reversal symmetry over the course of a phonon cycle. In contrast, circular excitation of two orthogonal components of a degenerate phonon mode produces a lattice trajectory with a definite handedness, as shown in Fig.~\ref{fig:mechanism}(a). The resulting circular dynamical Rashba coupling breaks time-reversal symmetry and reconstructs the electronic states into Floquet bands carrying finite Berry curvature, as shown in Fig.~\ref{fig:mechanism}(b). Reversing the phonon handedness reverses the induced Berry curvature and the associated Hall response.

To describe how the lattice motion produces this electronic response, we express the dynamical Rashba coupling in terms of the phonon-induced electric field. To leading order, this field is linear in the phonon normal-mode coordinate, $E_i^{\rm ph}(t)=\lambda_iQ_i(t)$, where $\lambda_i$ describes the electric field induced per unit displacement along phonon coordinate $Q_i$. The resulting phonon-induced spin-orbit Hamiltonian is~\cite{schlipf:2021}
\begin{equation}
H_{\rm SOC}(t)
=
\lambda [\mathbf{Q}(t)\times\mathbf{p}]
\cdot\mathbf{S}.
\label{eq:phonon_SOC}
\end{equation}
Driving two orthogonal components of the degenerate phonon with a
relative phase of $\pi/2$ produces a circular trajectory parametrized by
\begin{equation}
\mathbf{Q}(t)
=
Q_0(\cos\Omega t,\sin\Omega t,0),
\label{eq:circular_phonon}
\end{equation}
and Eq.~\eqref{eq:phonon_SOC} becomes
\begin{equation}
\begin{aligned}
H_{\rm SOC}(t)
={}&\lambda Q_0\bigl[
\cos(\Omega t)(p_yS_z-p_zS_y)\\
&+\sin(\Omega t)(p_zS_x-p_xS_z)
\bigr].
\end{aligned}
\label{eq:circular_SOC}
\end{equation}
The phonon-induced electric field follows the circular trajectory of $\mathbf{Q}(t)$. Reversing the phonon handedness reverses the temporal rotation of the coefficients coupling electronic momentum and spin in Eq.~\eqref{eq:circular_SOC}. We next examine how this periodic spin-orbit coupling modifies the electronic bands and generates an anomalous Hall response in a concrete material.

\begin{figure*}[t]
    \centering
    \includegraphics[width=0.85\textwidth]{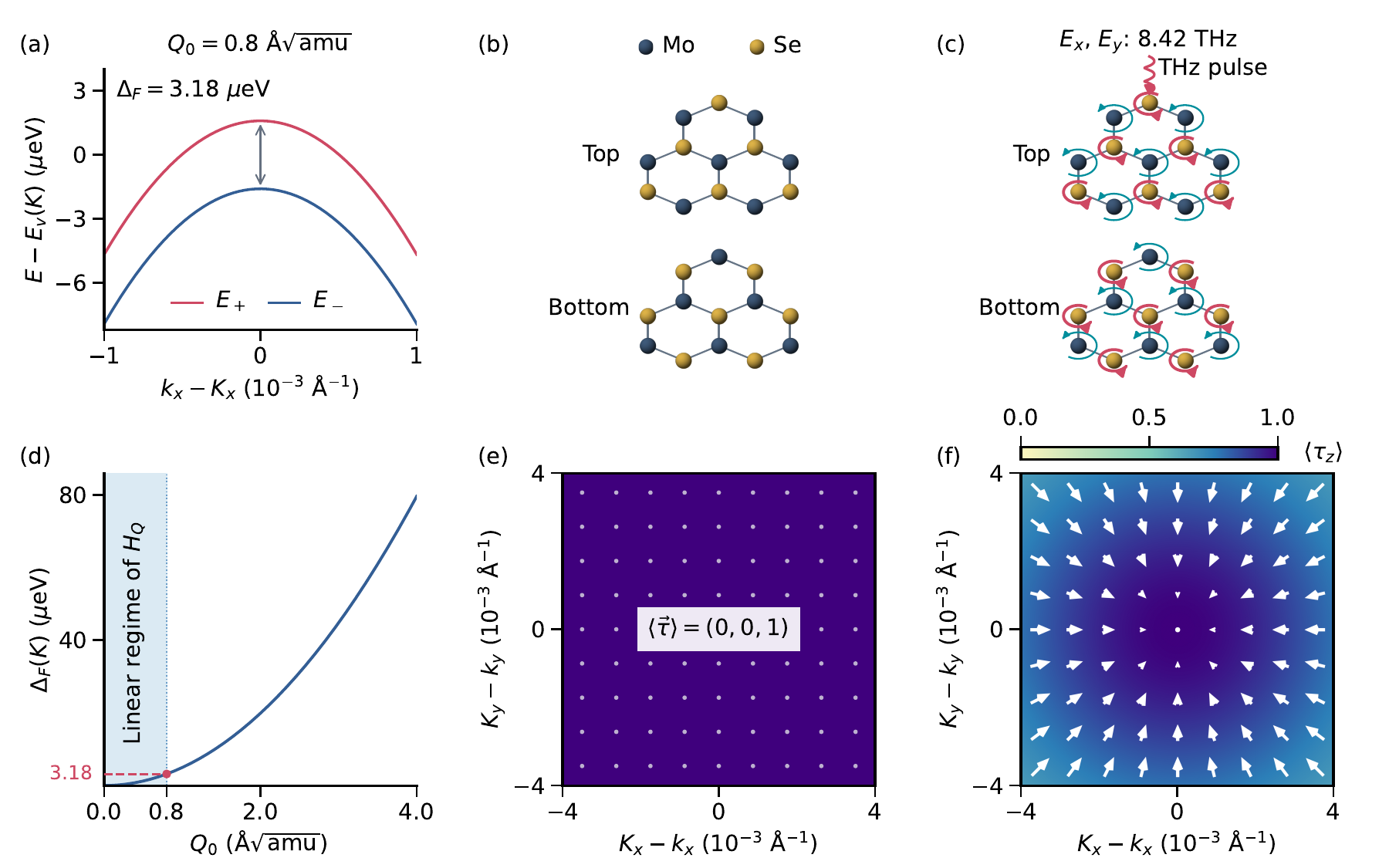}
\caption{\textbf{Circular dynamical Rashba coupling and Floquet band reconstruction in bilayer MoSe$_2$.}
(a) Floquet quasienergies from Eq.~\eqref{eq:floquet_bands} near $K$ for $Q_0=0.8\,\text{\AA}\sqrt{\mathrm{amu}}$, with valley-center gap $\Delta_F=3.18~\mu\mathrm{eV}$.
(b,c) Equilibrium structure and circular excitation of the $E_u$ phonon. The interlayer separation is enlarged for clarity in both panels. The orthogonal components are driven with a relative phase of $\pi/2$ at the phonon frequency $\Omega/(2\pi)=8.42$~THz.
(d) Phonon-amplitude-dependent Floquet gap from Eq.~\eqref{eq:floquet_gap}. Shading marks the validated linear electron-phonon coupling regime; the dot marks the amplitude used in (a).
(e,f) Equilibrium and Floquet-driven pseudospin textures from Eq.~\eqref{eq:floquet_H_compact}, illustrating the emergence of in-plane winding around the valley center and a finite out-of-plane component.}
    \label{fig:floquet}
\end{figure*}

For this purpose, we consider centrosymmetric 2H-stacked bilayer MoSe$_2$, a semiconductor with a measured electronic gap of $1.66$~eV~\cite{Li2022ElectronicGap}. Its two monolayers are coupled by van der Waals interactions~\cite{gong2013magnetoelectric,Lu2014,Zhang2014MoSe2} and arranged as shown in Fig.~\ref{fig:floquet}(b). Previous studies of chiral phonons in the related material monolayer MoS$_2$ predict chiral phonon magnetization~\cite{Zhang2023PhononMagnetization}
and report large effective phonon magnetic moments of approximately $2.5\mu_B$~\cite{Tang2024PhononMoment,Mustafa2025PhononMoment}. We calculate its equilibrium electronic structure and phonons
from first principles using density functional theory as implemented in
the Vienna \textit{Ab initio} Simulation Package (VASP)~\cite{kresse:1996,kresse2:1996}, including
spin-orbit coupling.
We identify a doubly degenerate in-plane $E_u$
phonon at $8.42$~THz. Since this mode is infrared active, its two orthogonal
components provide the inversion-breaking lattice coordinates required
for the circular excitation introduced in
Eq.~\eqref{eq:circular_phonon}. The equilibrium structure and the
circularly driven lattice are shown in Figs.~\ref{fig:floquet}(b) and \ref{fig:floquet}(c),
respectively.

Following Ref.~\cite{gong2013magnetoelectric}, we approximate the degenerate valence states near the $K$ valley by orbital-layer wavefunctions $u_1$ and $u_2$ with opposite out-of-plane spin projections,
\begin{equation}
|1\rangle=|u_1\rangle\otimes|\uparrow\rangle,
\qquad
|2\rangle=|u_2\rangle\otimes|\downarrow\rangle ,
\end{equation}
In this bilayer basis, $u_1$ and $u_2$ are formed primarily from layer-dependent combinations of the Mo $d_{x^2-y^2}$ and $d_{xy}$ orbitals. The two states have opposite layer polarizations, with finite weight on both layers due to interlayer hybridization~\cite{gong2013magnetoelectric}.
Pauli matrices $\boldsymbol{\tau}$ represent the pseudospin associated with the two basis states of this valence doublet, distinct from the physical spin $\mathbf{S}$ in Eq.~\eqref{eq:circular_SOC}.

We next project the phonon-induced spin-orbit coupling of
Eq.~\eqref{eq:phonon_SOC} onto this valence doublet subspace. The out-of-plane
momentum operator is represented by the matrix element $
P_z=\langle u_1|p_z|u_2\rangle ,
$ while, to leading order in momentum around $K$, the relevant in-plane
matrix elements can be written as $
p_x\simeq\hbar\alpha k_x,
 p_y\simeq\hbar\alpha k_y .
$
Because the two basis states have opposite spin projections, the terms containing $S_x$ and $S_y$ flip spin and couple the states through $P_z$, whereas the $S_z$ term is diagonal in this basis.
The resulting low-energy Hamiltonian is
\begin{equation}
H_Q(\mathbf{k})
=
\varepsilon_v(\mathbf{k})\tau_0
+
g(Q_y\tau_x-Q_x\tau_y)
+
h(Q_xk_y-Q_yk_x)\tau_z ,
\label{eq:projected_H}
\end{equation}
where
$g=\lambda|P_z|,\quad h=\lambda\hbar\alpha$,
and $\mathbf{k}$ is measured relative to the center of the $K$ valley. We approximate
the equilibrium valence-band dispersion by
$\varepsilon_v(\mathbf{k}) = \varepsilon_v(0)-ck^2$.
Here, $g$ controls the valley-center splitting, $h$ its momentum dependence, and $c$ the equilibrium band curvature.

We determine the parameters of Eq.~\eqref{eq:projected_H} using frozen-phonon calculations. For a static displacement
along one component of the $E_u$ mode, the splitting of the valence
doublet at the valley center is
$S(Q)=E_+(0,Q)-E_-(0,Q)=2gQ$.
The momentum dependence of the splitting provides an independent measure
of $h$. For a displacement $Q_x=Q$,
$S(k_y,Q) = 2|Q|\sqrt{g^2+h^2k_y^2}$.
We fit these expressions to the frozen-phonon band structures to obtain
$g$ and $h$, while $c$ is extracted from the equilibrium valence-band
dispersion. Details of the fitting procedure and the first-principles calculations using VASP~\cite{kresse:1996,kresse2:1996}, the PBEsol functional~\cite{perdew:2008}, and Phonopy~\cite{togo2023first} are provided in the Supplemental Material. Our calculations also determine the range of phonon amplitudes over which the leading-order electron-phonon coupling remains valid.

We next substitute the circular phonon trajectory of Eq.~\eqref{eq:circular_phonon} into Eq.~\eqref{eq:projected_H} to obtain a periodic Hamiltonian with period $T=2\pi/\Omega$. We consider the regime $
gQ_0,hQ_0k\ll\hbar\Omega\ll\Delta$,
where $\Delta$ is the energy separation from the modeled valence doublet to the nearest excluded band. At $K$, the equilibrium band calculation places the next lower valence doublet $217$~meV below the modeled doublet and the lowest conduction doublet $1.42$~eV above it, giving $\Delta_K\simeq217$~meV. At the valley center, the fitted coupling $g\simeq0.294~\mathrm{meV}/(\text{\AA}\sqrt{\mathrm{amu}})$ gives $gQ_0\simeq0.24~\mathrm{meV}$ at $Q_0=0.8\,\text{\AA}\sqrt{\mathrm{amu}}$, well below $\hbar\Omega\simeq34.8~\mathrm{meV}$. At $K$, the phonon energy is also about six times smaller than $\Delta_K$, supporting
the projection onto the valence doublet near $K$, while satisfying
$gQ_0,hQ_0k\ll\hbar\Omega$ ensures the validity of the high-frequency expansion within
this subspace.

Writing the time-dependent Hamiltonian as $ 
H(t)=H_0+H_{+1}e^{-i\Omega t}+H_{-1}e^{i\Omega t}
$, the effective Hamiltonian to first order in the Floquet expansion is given by $
H_F
=
H_0+
\frac{[H_{-1},H_{+1}]}{\hbar\Omega}$. Expanding the commutator, we obtain
\begin{equation}
H_F(\mathbf{k})
=
-ck^2\tau_0
-
\frac{Q_0^2}{\hbar\Omega}
\left[
gh(k_x\tau_x+k_y\tau_y)+g^2\tau_z
\right].
\label{eq:floquet_H_compact}
\end{equation}
This Hamiltonian contains a momentum-dependent in-plane pseudospin coupling and an out-of-plane mass term, both quadratic in the phonon amplitude. The mass term lifts the valence-doublet degeneracy at the valley center and enables finite Berry curvature.

\section{Results}
\label{sec:results}

We now evaluate the circular dynamical Rashba coupling and Floquet band reconstruction described by the Hamiltonian in Eq.~\eqref{eq:floquet_H_compact}, as summarized in Fig.~\ref{fig:floquet}. The corresponding Floquet quasienergies are
\begin{equation}
E_\pm(k)
=
-ck^2
\pm
\frac{Q_0^2}{\hbar\Omega}
\sqrt{g^2 h^2 k^2+g^4},
\label{eq:floquet_bands}
\end{equation}
and are shown near the $K$ valley in Fig.~\ref{fig:floquet}(a). At the valley center,
the Floquet mass lifts the valence-band degeneracy and opens a gap given by
\begin{equation}
\Delta_F
=
\frac{2g^2Q_0^2}{\hbar\Omega}.
\label{eq:floquet_gap}
\end{equation}
Figure~\ref{fig:floquet}(d) shows the quadratic dependence of the Floquet gap on the phonon amplitude predicted by Eq.~\eqref{eq:floquet_gap}. At $Q_0=0.8\,\text{\AA}\sqrt{\mathrm{amu}}$, the upper limit of the quantitatively validated linear electron-phonon coupling regime, the Floquet gap reaches $\Delta_F=3.18~\mu\mathrm{eV}$. Here, linear refers to the coupling $E_i^{\rm ph}=\lambda_iQ_i$, whereas the Floquet gap is quadratic in $Q_0$. These amplitudes yield atomic motions well below the Lindemann stability criterion of approximately $10\%$ of the interatomic distance, corresponding to $Q_0\simeq4\,\text{\AA}\sqrt{\mathrm{amu}}$. Beyond $0.8\,\text{\AA}\sqrt{\mathrm{amu}}$, the frozen-phonon splitting remains approximately linear in $Q_x$, but nonlinear corrections become important for $Q_y$ (Supplemental Material). Upon extrapolation of the leading-order expansion in Eq.~\eqref{eq:floquet_gap} to the Lindemann stability limit, the Floquet gap reaches  $80~\mu\mathrm{eV}$.

\begin{figure}[t]
\centering
\includegraphics[width=\columnwidth]{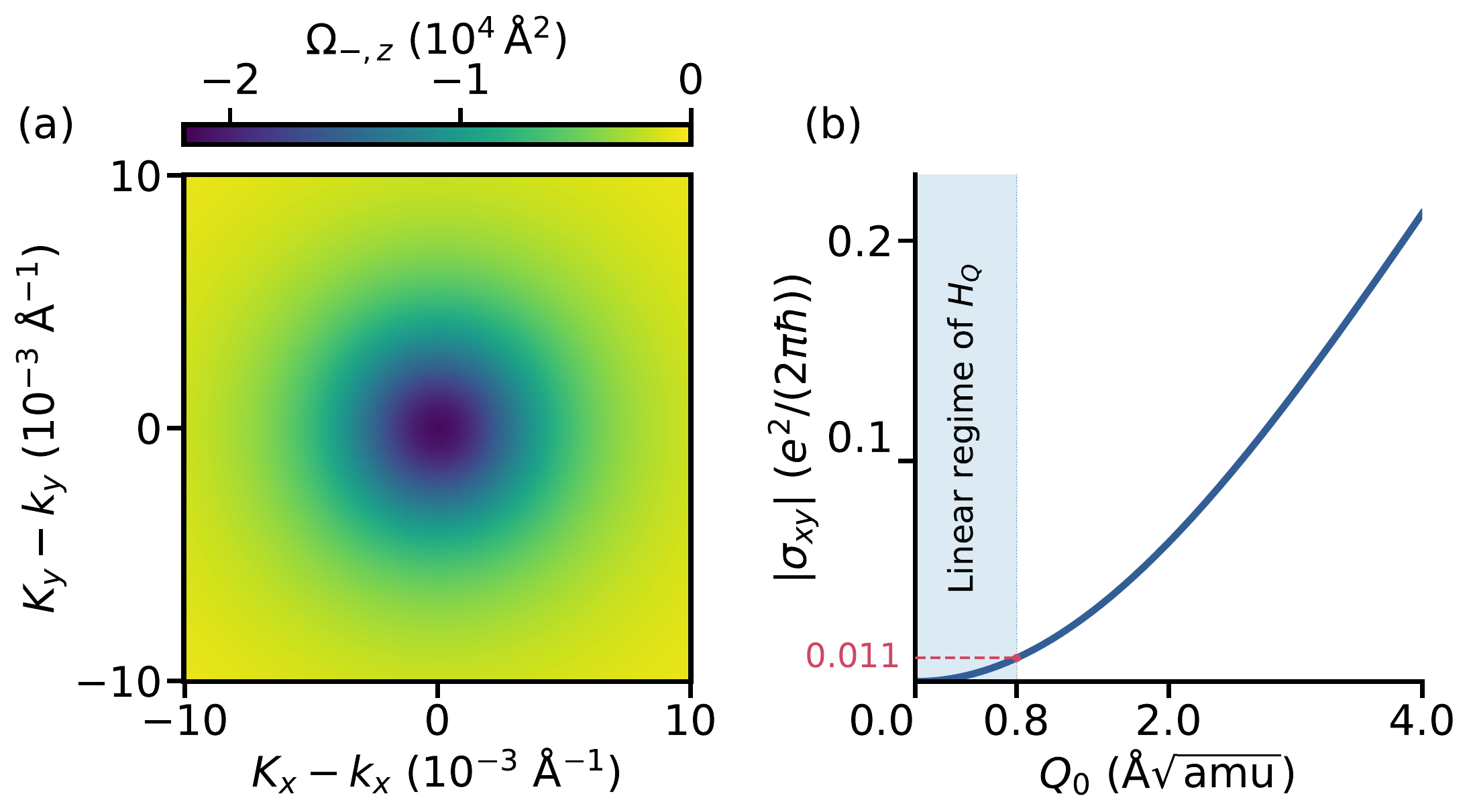}
    \caption{\textbf{Phonon-induced Berry curvature and anomalous Hall response.}
(a) Lower-band Berry curvature from Eq.~\eqref{eq:berry_curvature} near $K$ at $Q_0=0.8\,\text{\AA}\sqrt{\mathrm{amu}}$; reversing phonon handedness reverses its sign.
(b) Magnitude of the total Hall conductivity from Eq.~\eqref{eq:hall_vs_amplitude}, including both valleys, versus phonon amplitude. Shading marks the validated linear-coupling regime, with $|\sigma_{xy}|\simeq1.1\times10^{-2}e^2/(2\pi\hbar)$ at its upper edge. Upon extrapolation of the leading-order expansion to the Lindemann stability limit of $Q_0\simeq4\,\text{\AA}\sqrt{\mathrm{amu}}$, $|\sigma_{xy}|$ reaches approximately $0.21e^2/(2\pi\hbar)$.}
    \label{fig:hall}
\end{figure}

In Figs.~\ref{fig:floquet}(e) and (f), we compare the equilibrium and Floquet-driven pseudospin textures obtained from Eq.~\eqref{eq:floquet_H_compact}. Writing $H_F=d_0\tau_0+\mathbf{d}\cdot\boldsymbol{\tau}$ with $\mathbf{d}=-Q_0^2(ghk_x,ghk_y,g^2)/(\hbar\Omega)$, each Floquet-band pseudospin is aligned or antialigned with $\mathbf{d}$. Its in-plane component winds around the valley center, while both its in-plane and out-of-plane components are finite away from the valley center. The winding of $\hat{\mathbf{d}}=\mathbf{d}/|\mathbf{d}|$ determines the Berry curvature, which for the lower Floquet band is given by
\begin{equation}
\Omega_z^-(k)
=
\chi\frac{1}{2}
\frac{\xi}
{\left(k^2+\xi^2\right)^{3/2}},
\qquad
\xi=\frac{g}{h},
\label{eq:berry_curvature}
\end{equation}
Here, $\chi=\pm1$ specifies the Berry-curvature sign associated with the handedness of the chiral phonon. Equation~\eqref{eq:berry_curvature} has the characteristic form of
the Berry curvature of a two-dimensional massive Dirac cone, with
$\xi$ setting the characteristic momentum-space width of the
curvature distribution. In the low-energy model, where momentum is treated as a continuous variable near each valley, the Berry flux
associated with a single valley is
\begin{equation}
\frac{1}{2\pi}\int d^2k\,\Omega_z^-(k)=\frac{\chi}{2}.
\end{equation}
The
curvature is therefore concentrated around the valley center, as shown
in Fig.~\ref{fig:hall}(a), and reverses sign when the phonon chirality is reversed.

Importantly, the corresponding projection around the time-reversed
$K'$ valley gives a Berry curvature with the same sign for a fixed
phonon chirality. The two valleys therefore carry same-sign
Berry flux and contribute constructively
rather than destructively, resulting in a finite anomalous Hall response.
This is a direct consequence of the time-reversal-symmetry breaking
introduced by the chiral phonon, as in equilibrium the Berry curvatures of the two valleys have opposite signs.

The resulting anomalous Hall conductivity is~\cite{xiao:2010}
\begin{equation}
\sigma_{xy}
=
-\frac{e^2}{\hbar}
\sum_n
\int\frac{d^2k}{(2\pi)^2}
f_{n\mathbf{k}}\,
\Omega_{n,z}(\mathbf{k}).
\label{eq:hall_general}
\end{equation}
We set the Fermi energy to $E_F=E_-(0)$ and assume zero-temperature Fermi occupations of the effective Floquet bands. Under this assumption, the lower Floquet band is occupied throughout the relevant momentum range, whereas the upper band becomes occupied beyond a momentum $k_F$ determined by $E_+(k_F)=E_-(0)$. Since the two bands have opposite Berry curvatures, $\Omega_z^+=-\Omega_z^-$, their contributions cancel for $k>k_F$. The Hall response is therefore determined by the Berry curvature integrated over the disk $k<k_F$.

Using the Floquet dispersion in Eq.~\eqref{eq:floquet_bands} and the gap in Eq.~\eqref{eq:floquet_gap},
the occupation boundary satisfies
\begin{equation}
\sqrt{k_F^2+\xi^2}
=
\xi\left(1+\frac{h^2\Delta_F}{2cg^2}\right).
\end{equation}
Including both valleys, Eq.~\eqref{eq:hall_general} yields
\begin{equation}
\left|\sigma_{xy}\right|
=
\frac{e^2}{2\pi\hbar}
\frac{h^2\Delta_F}{2cg^2+h^2\Delta_F},
\label{eq:hall_vs_amplitude}
\end{equation}
which is the main result of our study.
A full derivation is provided in Sec.~III of the Supplemental
Material. Equation~\eqref{eq:hall_vs_amplitude} describes the dependence of the anomalous Hall conductivity on the chiral phonon amplitude, which we illustrate in Fig.~\ref{fig:hall}(b). An important feature of the leading-order
Floquet model is that the Berry-curvature profile in
Eq.~\eqref{eq:berry_curvature} is independent of $Q_0$ for any nonzero
drive amplitude, since the common factor $Q_0^2/(\hbar\Omega)$ multiplying
$\mathbf{d}(\mathbf{k})$ drops out of $\hat{\mathbf d}$. The increase of
the Hall conductivity with phonon amplitude therefore arises from the
growth of $k_F$, which enlarges the momentum-space region over which the
Berry-curvature contributions of the two Floquet bands do not cancel.
Within our first-principles validated
linear regime, at $Q_0=0.8\,\text{\AA}\sqrt{\mathrm{amu}}$, we obtain
$\left|\sigma_{xy}\right| \simeq
1.1\times10^{-2}\frac{e^2}{2\pi\hbar}$. Extrapolating to the Lindemann stability limit at an amplitude of $Q_0\simeq4\,\text{\AA}\sqrt{\mathrm{amu}}$ gives $|\sigma_{xy}|\sim0.21e^2/2\pi\hbar$. These numbers estimate a realistic magnitude of the anomalous Hall response, while an exact quantitative prediction at large amplitudes would require nonlinear electron-phonon couplings beyond the validated regime.

\section{Discussion}
\label{sec:discussion}

While our calculations focus on 2H-stacked bilayer MoSe$_2$, the mechanism
straightforwardly extends to other materials that combine strong spin-orbit coupling with degenerate infrared-active phonons that dynamically induce circular Rashba-type coupling. We anticipate that the effect can be measured in light-driven transport setups adapted to the terahertz regime~\cite{mciver:2020,lesko:2026}.

An experimental route toward the occupation regime considered here is to use top and bottom gates to introduce holes while keeping the perpendicular electric field small. In hBN-encapsulated, R-stacked bilayer WSe$_2$, optical reflectance measurements identified hole doping at zero applied perpendicular displacement field using graphite top and bottom gates~\cite{Li2026DualGateWSe2}. In 2H-stacked bilayer MoSe$_2$, a dual-gate device has been used to control carrier doping and the perpendicular field independently in optical measurements~\cite{Jasinski2025Quadrupolar}. These experiments motivate using a similar MoSe$_2$ device to tune the Fermi level near the valence-band edge while minimizing a static gate-induced splitting.

Phonon-mediated driving may also address practical limitations of direct
optical Floquet engineering. In light-driven graphene, the pump modifies both
the electronic bands and their occupations~\cite{mciver:2020}, and
photocarrier populations contribute substantially to the measured Hall
response~\cite{sato:2019}. In a gapped material, resonant excitation of an
infrared-active phonon instead offers a route to modulating the electronic
bands while suppressing direct interband
excitation~\cite{Forst2011Nonlinear, Disa2021}.
Moreover, coherent lattice motion can act as a long-lived Floquet
drive~\cite{Hubener2018Floquet}, and phonon-dressed Floquet--Bloch states have
been observed to outlive their light-driven counterparts by one to two orders
of magnitude~\cite{Tai2026PhononFloquet}, extending the
electronic dressing beyond the optical excitation window.
\section*{Data availability}

The data that support the findings of this article are openly available in Ref.~\cite{zenodoData2026}.

\begin{acknowledgments}

This work was supported by the Israel Science Foundation (ISF) Grant No. 1077/23 and 1916/23. This research was supported in part by grant no. NSF PHY-2309135 to the Kavli Institute for Theoretical Physics (KITP). O.Y. acknowledges support from the Rashba Excellence Award. C.P. acknowledges support from the Alexander Zaks scholarship. D.M.J. acknowledges support by the ERC Starting Grant CHIRALPHONONICS, No. 101166037.

\end{acknowledgments}

\bibliography{Bibliography}

@STRING( PRB = "Phys. Rev. B" )

@STRING( PRL = "Phys. Rev. Lett." )

@STRING( RMP = "Rev. Mod. Phys." )

@STRING{PR        = "Phys. Rev."}

@STRING{PRR       = "Phys. Rev. Research"}

@STRING{natphys    = "Nat. Phys."}

@STRING{compmat   = "Comput. Mat. Sci."}

@STRING{NL      = "Nano Lett."}

@article{juraschek:2017,
    author = {Juraschek, D. M. and Fechner, M. and Spaldin, N. A.},
    journal = PRL,
    pages = {054101},
    title = {{Ultrafast Structure Switching through Nonlinear Phononics}},
    doi = {10.1103/PhysRevLett.118.054101},
    volume = {118},
    year = {2017}
}

@article{Disa2021,
    author = {Disa, A. S. and Nova, T. F. and Cavalleri, A.},
    doi = {10.1038/s41567-021-01366-1},
    journal = natphys,
    pages = {1087--1092},
    title = {{Engineering crystal structures with light}},
    volume = {17},
    year = {2021}
}

@article{lindner:2011,
    author = {Lindner, N. H. and Refael, G. and Galitski, V.},
    journal = natphys,
    pages = {490--495},
    title = {{Floquet topological insulator in semiconductor quantum wells}},
    volume = {7},
    year = {2011},
  doi = {10.1038/nphys1926}
}

@article{perdew:2008,
    author = {Perdew, J. and Ruzsinszky, A. and Csonka, G. and Vydrov, O. and Scuseria, G. and Constantin, L. and Zhou, X. and Burke, K.},
    journal = PRL,
    pages = {136406},
    title = {{Restoring the Density-Gradient Expansion for Exchange in Solids and Surfaces}},
    doi = {10.1103/PhysRevLett.100.136406},
    volume = {100},
    year = {2008}
}

@article{nova:2017,
    author = {Nova, T. F. and Cartella, A. and Cantaluppi, A. and F{\"{o}}rst, M. and Bossini, D. and Mikhaylovskiy, R. V. and Kimel, A. V. and Merlin, R. and Cavalleri, A.},
    journal = natphys,
    pages = {132--137},
    title = {{An effective magnetic field from optically driven phonons}},
    doi = {10.1038/nphys3925},
    volume = {13},
    year = {2017}
}

@article{wang:2013,
    author = {Wang, Y. H. and Steinberg, H. and Jarillo-Herrero, P. and Gedik, N.},
    journal = {Science},
    title = {{Observation of Floquet-Bloch States on the Surface of a Topological Insulator}},
    volume = {342},
    doi = {10.1126/science.1239834},
    pages={453--457},
    year = {2013}
}

@article{kresse:1996,
    Author        = {G. Kresse and J. Furthm\"{u}ller},
    Journal       = compmat,
    Title         = {Efficiency of ab-initio total energy calculations for metals and semiconductors using a plane-wave basis set},
    Volume        = {6},
    doi = {10.1016/0927-0256(96)00008-0},
    Pages         = {15--50},
    Year          = {1996}
}

@article{kresse2:1996,
    Author        = {G. Kresse and J. Furthm\"{u}ller},
    Journal       = PRB,
    Title         = {Efficient iterative schemes for ab initio total-energy calculations using a plane-wave basis set},
    Volume        = {54},
    doi = {10.1103/PhysRevB.54.11169},
    Pages         = {11169},
    Year          = {1996}
}

@article{Xiao2021,
    author = {Xiao, C. and Ren, Y. and Xiong, B.},
    doi = {10.1103/PhysRevB.103.115432},
    journal = PRB,
    pages = {115432},
    title = {{Adiabatically induced orbital magnetization}},
    volume = {103},
    year = {2021}
}

@article{Juraschek2025,
author={Juraschek, Dominik M.
and Geilhufe, R. Matthias
and Zhu, Hanyu
and Basini, Martina
and Baum, Peter
and Baydin, Andrey
and Chaudhary, Swati
and Fechner, Michael
and Flebus, Benedetta
and Grissonnanche, Gael
and Kirilyuk, Andrei I.
and Lemeshko, Mikhail
and Maehrlein, Sebastian F.
and Mignolet, Maxime
and Murakami, Shuichi
and Niu, Qian
and Nowak, Ulrich
and Romao, Carl P.
and Rostami, Habib
and Satoh, Takuya
and Spaldin, Nicola A.
and Ueda, Hiroki
and Zhang, Lifa},
title={Chiral phonons},
journal=natphys,
year={2025},
volume={21},
pages={1532–1540},
doi={10.1038/s41567-025-03001-9}
}

@misc{zenodoData2026,
  title = {{Original code available at Zenodo}},
  author = {Yaniv, Omer and Chaudhary, Swati and Paiva, Carolina and Juraschek, Dominik M.},
  year = 2026
}

@article{karplus:1954,
    author = {Karplus, R. and Luttinger, J. M.},
    journal = PR,
    pages = {1154},
    title = {{Hall Effect in Ferromagnetics}},
    doi = {10.1103/PhysRev.95.1154},
    volume = {95},
    year = {1954}
}

@article{haldane:2004,
    author = {Haldane, F. D. M.},
    journal = PRL,
    pages = {206602},
    title = {{Berry Curvature on the Fermi Surface: Anomalous Hall Effect as a Topological Fermi-Liquid Property}},
    doi = {10.1103/PhysRevLett.93.206602},
    volume = {93},
    year = {2004}
}

@article{nagaosa:2010,
    author = {Nagaosa, N. and Sinova, J. and Onoda, S. and MacDonald, A. H. and Ong, N. P.},
    journal = RMP,
    pages = {1539},
    title = {{Anomalous Hall Effect}},
    doi = {10.1103/RevModPhys.82.1539},
    volume = {82},
    year = {2010}
}

@article{sundaram:1999,
    author = {Sundaram, G. and Niu, Q.},
    journal = PRB,
    pages = {14915},
    title = {{Wave-Packet Dynamics in Slowly Perturbed Crystals: Gradient Corrections and Berry-Phase Effects}},
    doi = {10.1103/PhysRevB.59.14915},
    volume = {59},
    year = {1999}
}

@article{xiao:2010,
    author = {Xiao, D. and Chang, M.-C. and Niu, Q.},
    journal = RMP,
    pages = {1959},
    title = {{Berry Phase Effects on Electronic Properties}},
    doi = {10.1103/RevModPhys.82.1959},
    volume = {82},
    year = {2010}
}

@article{oka:2009,
    author = {Oka, T. and Aoki, H.},
    journal = PRB,
    pages = {081406},
    title = {{Photovoltaic Hall Effect in Graphene}},
    doi = {10.1103/PhysRevB.79.081406},
    volume = {79},
    year = {2009}
}

@article{rudner:2020,
    author = {Rudner, M. S. and Lindner, N. H.},
    journal = {Nat. Rev. Phys.},
    pages = {229},
    title = {{Band Structure Engineering and Non-Equilibrium Dynamics in Floquet Topological Insulators}},
    doi = {10.1038/s42254-020-0170-z},
    volume = {2},
    year = {2020}
}

@article{mciver:2020,
    author = {McIver, J. W. and Schulte, B. and Stein, F.-U. and Matsuyama, T. and Jotzu, G. and Meier, G. and Cavalleri, A.},
    journal = natphys,
    pages = {38},
    title = {{Light-Induced Anomalous Hall Effect in Graphene}},
    doi = {10.1038/s41567-019-0698-y},
    volume = {16},
    year = {2020}
}

@article{merboldt:2025,
    author = {Merboldt, M. and Sch{\"u}ler, M. and Schmitt, D. and Bange, J. P. and Bennecke, W. and Gadge, K. and Pierz, K. and Schumacher, H. W. and Momeni, D. and Steil, D. and Manmana, S. R. and Sentef, M. A. and Reutzel, M. and Mathias, S.},
    journal = natphys,
    pages = {1093},
    title = {{Observation of Floquet States in Graphene}},
    doi = {10.1038/s41567-025-02889-7},
    volume = {21},
    year = {2025}
}

@article{lesko:2026,
    author = {Lesko, D. M. B. and Weitz, T. and Wittigschlager, S. and Li, W. and Heide, C. and Neufeld, O. and Hommelhoff, P.},
    journal = natphys,
    title = {{Optical Control of Electrons in a Floquet Topological Insulator}},
    doi = {10.1038/s41567-026-03429-7},
    year = {2026}
}

@article{sato:2019,
    author = {Sato, S. A. and McIver, J. W. and Nuske, M. and Tang, P. and Jotzu, G. and Schulte, B. and H{\"u}bener, H. and De Giovannini, U. and Mathey, L. and Sentef, M. A. and Cavalleri, A. and Rubio, A.},
    journal = PRB,
    pages = {214302},
    title = {{Microscopic Theory for the Light-Induced Anomalous Hall Effect in Graphene}},
    doi = {10.1103/PhysRevB.99.214302},
    volume = {99},
    year = {2019}
}

@article{Lu2014,
    author = {Lu, X. and Utama, M. I. B. and Lin, J. and Gong, X. and Zhang, J. and Zhao, Y. and Pantelides, S. T. and Wang, J. and Dong, Z. and Liu, Z. and Zhou, W. and Xiong, Q.},
    journal = NL,
    pages = {2419},
    title = {{Large-Area Synthesis of Monolayer and Few-Layer MoSe$_2$ Films on SiO$_2$ Substrates}},
    doi = {10.1021/nl5000906},
    volume = {14},
    year = {2014}
}

@article{juraschek:2019,
    author = {Juraschek, D. M. and Spaldin, N. A.},
    journal = {Phys. Rev. Materials},
    pages = {064405},
    title = {{Orbital Magnetic Moments of Phonons}},
    doi = {10.1103/PhysRevMaterials.3.064405},
    volume = {3},
    year = {2019}
}

@article{juraschek:2022,
    author = {Juraschek, D. M. and Neuman, T. and Narang, P.},
    journal = PRR,
    pages = {013129},
    title = {{Giant Effective Magnetic Fields from Optically Driven Chiral Phonons in $4f$ Paramagnets}},
    doi = {10.1103/PhysRevResearch.4.013129},
    volume = {4},
    year = {2022}
}

@article{luo:2023,
    author = {Luo, J. and Lin, T. and Zhang, J. and Chen, X. and Blackert, E. R. and Xu, R. and Yakobson, B. I. and Zhu, H.},
    journal = {Science},
    pages = {698},
    title = {{Large Effective Magnetic Fields from Chiral Phonons in Rare-Earth Halides}},
    doi = {10.1126/science.adi9601},
    volume = {382},
    year = {2023}
}

@article{schlipf:2021,
    author = {Schlipf, M. and Giustino, F.},
    journal = PRL,
    pages = {237601},
    title = {{Dynamic Rashba-Dresselhaus Effect}},
    doi = {10.1103/PhysRevLett.127.237601},
    volume = {127},
    year = {2021}
}

@article{gong2013magnetoelectric,
  title={Magnetoelectric effects and valley-controlled spin quantum gates in transition metal dichalcogenide bilayers},
  author={Gong, Zhirui and Liu, Gui-Bin and Yu, Hongyi and Xiao, Di and Cui, Xiaodong and Xu, Xiaodong and Yao, Wang},
  journal={Nat. Commun.},
  volume={4},
  number={1},
  pages={2053},
  year={2013},
  doi = {10.1038/ncomms3053},
  publisher={Nature Publishing Group UK London}
}

@article{Juraschek2017Dynamical,
  author = {Dominik M. Juraschek and Michael Fechner and Alexander V. Balatsky and Nicola A. Spaldin},
  title = {{Dynamical multiferroicity}},
  journal = {Phys. Rev. Materials},
  volume = {1},
  pages = {014401},
  year = {2017},
  doi = {10.1103/PhysRevMaterials.1.014401},
  url = {https://doi.org/10.1103/PhysRevMaterials.1.014401}
}

@article{Shabala2024Inverse,
  author = {Natalia Shabala and R. Matthias Geilhufe},
  title = {{Phonon Inverse Faraday Effect from Electron-Phonon Coupling}},
  journal = {Phys. Rev. Lett.},
  volume = {133},
  pages = {266702},
  year = {2024},
  doi = {10.1103/PhysRevLett.133.266702},
  url = {https://doi.org/10.1103/PhysRevLett.133.266702}
}

@article{Klebl2025Geometry,
  author = {Lennart Klebl and Arne Schobert and Martin Eckstein and Giorgio Sangiovanni and Alexander V. Balatsky and Tim O. Wehling},
  title = {{Ultrafast Pseudomagnetic Fields from Electron-Nuclear Quantum Geometry}},
  journal = {Phys. Rev. Lett.},
  volume = {134},
  pages = {016705},
  year = {2025},
  doi = {10.1103/PhysRevLett.134.016705},
  url = {https://doi.org/10.1103/PhysRevLett.134.016705}
}

@article{Basini2024Multiferroicity,
  author = {M. Basini and M. Pancaldi and B. Wehinger and M. Udina and V. Unikandanunni and T. Tadano and M. C. Hoffmann and A. V. Balatsky and S. Bonetti},
  title = {{Terahertz electric-field-driven dynamical multiferroicity in {SrTiO$_3$}}},
  journal = {Nature},
  volume = {628},
  pages = {534--539},
  year = {2024},
  doi = {10.1038/s41586-024-07175-9},
  url = {https://doi.org/10.1038/s41586-024-07175-9}
}

@article{Hubener2018Floquet,
  author = {Hannes H{\"u}bener and Umberto De Giovannini and Angel Rubio},
  title = {{Phonon Driven Floquet Matter}},
  journal = {Nano Lett.},
  volume = {18},
  pages = {1535--1542},
  year = {2018},
  doi = {10.1021/acs.nanolett.7b05391},
  url = {https://doi.org/10.1021/acs.nanolett.7b05391}
}

@article{Shin2018SpinFloquet,
  author = {Dongbin Shin and Hannes H{\"u}bener and Umberto De Giovannini and Hosub Jin and Angel Rubio and Noejung Park},
  title = {{Phonon-driven spin-Floquet magneto-valleytronics in {MoS$_2$}}},
  journal = {Nat. Commun.},
  volume = {9},
  pages = {638},
  year = {2018},
  doi = {10.1038/s41467-018-02918-5},
  url = {https://doi.org/10.1038/s41467-018-02918-5}
}

@article{Forst2011Nonlinear,
  author = {M. F{\"o}rst and C. Manzoni and S. Kaiser and Y. Tomioka and Y. Tokura and R. Merlin and A. Cavalleri},
  title = {{Nonlinear phononics as an ultrafast route to lattice control}},
  journal = {Nat. Phys.},
  volume = {7},
  pages = {854--856},
  year = {2011},
  doi = {10.1038/nphys2055},
  url = {https://doi.org/10.1038/nphys2055}
}

@article{Subedi2014Nonlinear,
  author = {Alaska Subedi and Andrea Cavalleri and Antoine Georges},
  title = {{Theory of nonlinear phononics for coherent light control of solids}},
  journal = {Phys. Rev. B},
  volume = {89},
  pages = {220301},
  year = {2014},
  doi = {10.1103/PhysRevB.89.220301},
  url = {https://doi.org/10.1103/PhysRevB.89.220301}
}

@article{Zhang2015Chiral,
  author = {Lifa Zhang and Qian Niu},
  title = {{Chiral Phonons at High-Symmetry Points in Monolayer Hexagonal Lattices}},
  journal = {Phys. Rev. Lett.},
  volume = {115},
  pages = {115502},
  year = {2015},
  doi = {10.1103/PhysRevLett.115.115502},
  url = {https://doi.org/10.1103/PhysRevLett.115.115502}
}

@article{Zhu2018Chiral,
  author = {Hanyu Zhu and Jun Yi and Ming-Yang Li and Jun Xiao and Lifa Zhang and Chih-Wen Yang and Robert A. Kaindl and Lain-Jong Li and Yuan Wang and Xiang Zhang},
  title = {{Observation of chiral phonons}},
  journal = {Science},
  volume = {359},
  pages = {579--582},
  year = {2018},
  doi = {10.1126/science.aar2711},
  url = {https://doi.org/10.1126/science.aar2711}
}

@article{Shirley1965Floquet,
  author = {Jon H. Shirley},
  title = {{Solution of the Schr{\"o}dinger Equation with a Hamiltonian Periodic in Time}},
  journal = {Phys. Rev.},
  volume = {138},
  pages = {B979--B987},
  year = {1965},
  doi = {10.1103/PhysRev.138.B979},
  url = {https://doi.org/10.1103/PhysRev.138.B979}
}

@article{Sambe1973Quasienergies,
  author = {Hideo Sambe},
  title = {{Steady States and Quasienergies of a Quantum-Mechanical System in an Oscillating Field}},
  journal = {Phys. Rev. A},
  volume = {7},
  pages = {2203--2213},
  year = {1973},
  doi = {10.1103/PhysRevA.7.2203},
  url = {https://doi.org/10.1103/PhysRevA.7.2203}
}

@article{Bukov2015HighFrequency,
  author = {Marin Bukov and Luca D{\textquotesingle}Alessio and Anatoli Polkovnikov},
  title = {{Universal high-frequency behavior of periodically driven systems: from dynamical stabilization to Floquet engineering}},
  journal = {Adv. Phys.},
  volume = {64},
  pages = {139--226},
  year = {2015},
  doi = {10.1080/00018732.2015.1055918},
  url = {https://doi.org/10.1080/00018732.2015.1055918}
}

@article{Jungwirth2002AHE,
  author = {T. Jungwirth and Qian Niu and A. H. MacDonald},
  title = {{Anomalous Hall Effect in Ferromagnetic Semiconductors}},
  journal = {Phys. Rev. Lett.},
  volume = {88},
  pages = {207208},
  year = {2002},
  doi = {10.1103/PhysRevLett.88.207208},
  url = {https://doi.org/10.1103/PhysRevLett.88.207208}
}

@article{Fang2003Monopoles,
  author = {Zhong Fang and Naoto Nagaosa and Kei S. Takahashi and Atsushi Asamitsu and Roland Mathieu and Takeshi Ogasawara and Hiroyuki Yamada and Masashi Kawasaki and Yoshinori Tokura and Kiyoyuki Terakura},
  title = {{The Anomalous Hall Effect and Magnetic Monopoles in Momentum Space}},
  journal = {Science},
  volume = {302},
  pages = {92--95},
  year = {2003},
  doi = {10.1126/science.1089408},
  url = {https://doi.org/10.1126/science.1089408}
}

@article{Yao2004AHE,
  author = {Yugui Yao and Leonard Kleinman and A. H. MacDonald and Jairo Sinova and T. Jungwirth and Ding-sheng Wang and Enge Wang and Qian Niu},
  title = {{First Principles Calculation of Anomalous Hall Conductivity in Ferromagnetic bcc {Fe}}},
  journal = {Phys. Rev. Lett.},
  volume = {92},
  pages = {037204},
  year = {2004},
  doi = {10.1103/PhysRevLett.92.037204},
  url = {https://doi.org/10.1103/PhysRevLett.92.037204}
}

@article{Kitagawa2011Transport,
  author = {Takuya Kitagawa and Takashi Oka and Arne Brataas and Liang Fu and Eugene Demler},
  title = {{Transport properties of nonequilibrium systems under the application of light: Photoinduced quantum Hall insulators without Landau levels}},
  journal = {Phys. Rev. B},
  volume = {84},
  pages = {235108},
  year = {2011},
  doi = {10.1103/PhysRevB.84.235108},
  url = {https://doi.org/10.1103/PhysRevB.84.235108}
}

@article{PerezPiskunow2014Floquet,
  author = {P. M. Perez-Piskunow and Gonzalo Usaj and C. A. Balseiro and L. E. F. Foa Torres},
  title = {{Floquet chiral edge states in graphene}},
  journal = {Phys. Rev. B},
  volume = {89},
  pages = {121401},
  year = {2014},
  doi = {10.1103/PhysRevB.89.121401},
  url = {https://doi.org/10.1103/PhysRevB.89.121401}
}

@article{Zhang2014AngularMomentum,
  author = {Lifa Zhang and Qian Niu},
  title = {{Angular Momentum of Phonons and the Einstein--de Haas Effect}},
  journal = {Phys. Rev. Lett.},
  volume = {112},
  pages = {085503},
  year = {2014},
  doi = {10.1103/PhysRevLett.112.085503},
  url = {https://doi.org/10.1103/PhysRevLett.112.085503}
}

@article{Ren2021TopologicalMagnetization,
  author = {Yafei Ren and Cong Xiao and Daniyar Saparov and Qian Niu},
  title = {{Phonon Magnetic Moment from Electronic Topological Magnetization}},
  journal = {Phys. Rev. Lett.},
  volume = {127},
  pages = {186403},
  year = {2021},
  doi = {10.1103/PhysRevLett.127.186403},
  url = {https://doi.org/10.1103/PhysRevLett.127.186403}
}

@article{Geilhufe2021Magnetism,
  author = {R. Matthias Geilhufe and Vladimir Juri{\v c}i{\'c} and Stefano Bonetti and Jian-Xin Zhu and Alexander V. Balatsky},
  title = {{Dynamically induced magnetism in {KTaO$_3$}}},
  journal = {Phys. Rev. Research},
  volume = {3},
  pages = {L022011},
  year = {2021},
  doi = {10.1103/PhysRevResearch.3.L022011},
  url = {https://doi.org/10.1103/PhysRevResearch.3.L022011}
}

@article{Iadecola2013DrivenGraphene,
  author  = {Iadecola, Thomas and Campbell, David and Chamon, Claudio
             and Hou, Chang-Yu and Jackiw, Roman and Pi, So-Young
             and Kusminskiy, Silvia Viola},
  title   = {Materials Design from Nonequilibrium Steady States:
             Driven Graphene as a Tunable Semiconductor with
             Topological Properties},
  journal = {Phys. Rev. Lett.},
  volume  = {110},
  pages   = {176603},
  year    = {2013},
  doi     = {10.1103/PhysRevLett.110.176603}
}

@article{Chaudhary2020PhononFloquet,
  author  = {Chaudhary, Swati and Haim, Arbel and Peng, Yang
             and Refael, Gil},
  title   = {Phonon-induced {Floquet} topological phases protected
             by space-time symmetries},
  journal = {Phys. Rev. Research},
  volume  = {2},
  pages   = {043431},
  year    = {2020},
  doi     = {10.1103/PhysRevResearch.2.043431}
}

@article{Yang2024Acoustoelectric,
  author  = {Yang, Christopher and Hunt, Will and Refael, Gil
             and Esin, Iliya},
  title   = {Quantized Acoustoelectric {Floquet} Effect
             in Quantum Nanowires},
  journal = {Phys. Rev. Lett.},
  volume  = {133},
  pages   = {226301},
  year    = {2024},
  doi     = {10.1103/PhysRevLett.133.226301}
}

@misc{Yao2026PhononFloquet,
  author        = {Yao, Dapeng and Zhang, Tiantian and Oka, Takashi
                   and Yokoyama, Takehito},
  title         = {Magnetism and Topology from Circularly Polarized
                   Phonon {Floquet} Engineering},
  year          = {2026},
  eprint        = {2606.10854},
  archivePrefix = {arXiv},
  primaryClass  = {cond-mat.mes-hall}
}

@misc{Tai2026PhononFloquet,
  author        = {Tai, Yu-Chan and Luo, Chih-Wei and Takagi, Noriaki
                   and Ishida, Hiroshi and Lin, Chun-Liang
                   and Arafune, Ryuichi},
  title         = {Phonon-driven {Floquet} engineering
                   for electronic states},
  year          = {2026},
  eprint        = {2606.30065},
  archivePrefix = {arXiv},
  primaryClass  = {cond-mat.mtrl-sci}
}

@article{Zhang2023PhononMagnetization,
  author  = {Zhang, Shengjie and Pei, Yufei and Hu, Shiqi
             and Wu, Na and Chen, Da-Qiang and Lian, Chao
             and Meng, Sheng},
  title   = {Light-Induced Phonon-Mediated Magnetization
             in Monolayer {MoS$_2$}},
  journal = {Chin. Phys. Lett.},
  volume  = {40},
  number  = {7},
  pages   = {077502},
  year    = {2023},
  doi     = {10.1088/0256-307X/40/7/077502}
}

@article{Tang2024PhononMoment,
  author  = {Tang, Chunli and Ye, Gaihua and Nnokwe, Cynthia
             and Fang, Mengqi and Xiang, Li
             and Mahjouri-Samani, Masoud and Smirnov, Dmitry
             and Yang, Eui-Hyeok and Wang, Tingting
             and Zhang, Lifa and He, Rui and Jin, Wencan},
  title   = {Exciton-activated effective phonon magnetic
             moment in monolayer {MoS$_2$}},
  journal = {Phys. Rev. B},
  volume  = {109},
  number  = {15},
  pages   = {155426},
  year    = {2024},
  doi     = {10.1103/PhysRevB.109.155426}
}

@article{Mustafa2025PhononMoment,
  author  = {Mustafa, Hussam and Nnokwe, Cynthia and Ye, Gaihua
             and Fang, Mengqi and Chaudhary, Swati and Yan, Jia-An
             and Wu, Kai and Cunningham, Connor J.
             and Hemesath, Colin M. and Stollenwerk, Andrew James
             and Shand, Paul M. and Yang, Eui-Hyeok
             and Fiete, Gregory A. and He, Rui and Jin, Wencan},
  title   = {Origin of Large Effective Phonon Magnetic Moments
             in Monolayer {MoS$_2$}},
  journal = {ACS Nano},
  volume  = {19},
  number  = {11},
  pages   = {11241--11248},
  year    = {2025},
  doi     = {10.1021/acsnano.4c18906}
}

@article{Biggs2026ChiralPhononPolaritons,
  author  = {Biggs, Megan F. and Ho, Sin-Hang Enoch
             and Alejandro, Aldair and Lutz, Matthew J.
             and Moss, Clayton D. and Johnson, Jeremy A.},
  title   = {Ultrafast {Faraday} rotation probe of chiral
             phonon-polaritons in {LiNbO$_3$}},
  journal = {Sci. Adv.},
  volume  = {12},
  number  = {36},
  pages   = {eaec8970},
  year    = {2026},
  doi     = {10.1126/sciadv.aec8970}
}

@article{Davies2024UltrafastBarnett,
  author  = {Davies, C. S. and Fennema, F. G. N.
             and Tsukamoto, A. and Razdolski, I.
             and Kimel, A. V. and Kirilyuk, A.},
  title   = {Phononic switching of magnetization by the
             ultrafast {Barnett} effect},
  journal = {Nature},
  volume  = {628},
  number  = {8008},
  pages   = {540--544},
  year    = {2024},
  doi     = {10.1038/s41586-024-07200-x}
}

@article{Wang2025PhononMoment,
 author={Wang, Fuyi and Liu, Xinqi and Sun, Hong and Wang, Huaiqiang and Murakami, Shuichi and Zhang, Lifa and Zhang, Haijun and Xing, Dingyu},
 title={{Ab Initio Theory of Phonon Magnetic Moment Induced by Electron-Phonon Coupling in Magnetic Materials}},
 journal=PRL, volume={135}, pages={256701}, year={2025}, doi={10.1103/tpjd-dh1m}
}

@article{Yao2026Orbital,
 author={Yao, Dapeng and Go, Dongwook and Mokrousov, Yuriy and Murakami, Shuichi},
 title={{Dynamical Orbital Angular Momentum Induced by Circularly Polarized Phonons}},
 journal=PRL, volume={137}, pages={066301}, year={2026}, doi={10.1103/mnth-3g9n}
}

@article{Sato2026Orbital,
 author={Sato, Tetsuya and Kato, Takeo and Manchon, Aurelien},
 title={{Orbital Accumulation Induced by Chiral Phonons}},
 journal=PRL, volume={137}, pages={066302}, year={2026}, doi={10.1103/2yqn-cyzg}
}

@misc{Shabala2025Axial,
 author={Shabala, Natalia and Tietjen, Finja and Geilhufe, R. Matthias},
 title={{Axial phono-magnetic effects}}, year={2025}, eprint={2511.03329}, archivePrefix={arXiv}, primaryClass={cond-mat.mtrl-sci}
}

@misc{Zalewski2026CrystalField,
 author={Zalewski, T. and Mrudul, M. S. and Lee, Y. and Weissenhofer, M. and Boris, A. V. and Oppeneer, P. M. and Kirilyuk, A. and Davies, C. S.},
 title={{A crystal-field route to THz-driven magnetization}}, year={2026}, eprint={2607.08519}, archivePrefix={arXiv}, primaryClass={cond-mat.mtrl-sci}
}

@misc{Shabala2026Unifying,
 author={Shabala, Natalia and Tietjen, Finja and Liljegren, Ylva and Geilhufe, R. Matthias},
 title={{Unifying microscopic theories for the phono-magnetic effect}}, year={2026}, eprint={2608.04754}, archivePrefix={arXiv}, primaryClass={cond-mat.mtrl-sci}
}

@misc{Huang2026Axial,
 author={Huang, Chuankun and Milner, Alexander and Luo, Jiaming and Ye, Jianbo and Ye, Gaihua and Zhang, Junjie and Nnokwe, Cynthia and Yakobson, Boris I. and He, Rui and Milner, Valery and Zhu, Hanyu},
 title={{Ultrafast Magnetization Induced from Raman-Active Axial Chiral Phonons}}, year={2026}, eprint={2609.14179}, archivePrefix={arXiv}, primaryClass={cond-mat.mtrl-sci}
}

@misc{Bilyk2026Domains,
 author={Bilyk, Vladislav and Khokhlov, Nikolai and Radovskaia, Viktoriia and Kim, Peter and Diona, Pietro and Kaushik, Ravi and Maranzana, Luca and Hayashida, Takeshi and Artyukhin, Sergey and Teo, Edwin Hang Tong and Chaturvedi, Apoorva and Davies, Carl S. and Kirilyuk, Andrei and Kimel, Alexey and Afanasiev, Dmytro},
 title={{Deterministic control of antiferromagnetic domain walls by circular phonons}}, year={2026}, eprint={2609.23563}, archivePrefix={arXiv}, primaryClass={cond-mat.mtrl-sci}
}

@misc{Allington2026Spin,
 author={Allington, Clifford J. and Lutz, Matthew J. and Ho, Sin-Hang Enoch and Graf, Fabian and Basini, Martina and Ueda, Hiroki and Grimes, Michael and Elliott, Alexander B. and Biggs, Megan F. and Finn, Ravi and Huang, Shih-Wen and Henstridge, Merideth A. and Hoffmann, Matthias C. and Sato, Takahiro and Alonso-Mori, Roberto and Zhu, Diling and Nguyen, Quynh L. and Esposito, Vincent and Babicz Jr., Jeffrey T. and Skoropata, Elizabeth and Liu, Biaolong and Paris, Eugenio and Romaguera Camps, Arnau and Razzoli, Elia and Mankowsky, Roman and Capotondi, Flavio and Jaouen, Nicolas and Pergolesi, Daniele and Radovic, Milan and Savoini, Matteo and Johnson, Steven L. and Johnson, Jeremy A. and Staub, Urs},
 title={{Disentangling spin polarization from driven circular ionic motion in EuTiO$_3$}}, year={2026}, eprint={2609.28214}, archivePrefix={arXiv}, primaryClass={cond-mat.mtrl-sci}
}

@misc{Choi2026Ionic,
 author={Choi, In Hyeok and Urazhdin, Sergei and Wong, Man Tou and Liu, Zi-Jie and Nelson, Keith A.},
 title={{THz-Driven Quantum Ionic Magnetism in a Quantum Paraelectric SrTiO$_3$}}, year={2026}, eprint={2609.27347}, archivePrefix={arXiv}, primaryClass={cond-mat.other}
}

@article{Zhang2014MoSe2,
 author={Zhang, Yi and Chang, Tay-Rong and Zhou, Bo and Cui, Yong-Tao and Yan, Hao and Liu, Zhongkai and Schmitt, Felix and Lee, James and Moore, Rob and Chen, Yulin and Lin, Hsin and Jeng, Horng-Tay and Mo, Sung-Kwan and Hussain, Zahid and Bansil, Arun and Shen, Zhi-Xun},
 title={{Direct observation of the transition from indirect to direct bandgap in atomically thin epitaxial MoSe$_2$}},
 journal={Nat. Nanotechnol.}, volume={9}, pages={111--115}, year={2014}, doi={10.1038/nnano.2013.277}
}

@article{Li2022ElectronicGap,
  author = {Li, Dian and Wang, Xiong and Mo, Xiaoyong and Tse, Edmund C. M. and Cui, Xiaodong},
  title = {Electronic gap characterization at mesoscopic scale via scanning probe microscopy under ambient conditions},
  journal = {Nat. Commun.},
  volume = {13},
  pages = {4648},
  year = {2022},
  doi = {10.1038/s41467-022-32439-1}
}

@article{Jasinski2025Quadrupolar,
  author = {Jasi{\'n}ski, Jakub and Hagel, Joakim and Brem, Samuel and Wietek, Edith and Taniguchi, Takashi and Watanabe, Kenji and Chernikov, Alexey and Bruyant, Nicolas and Dyksik, Mateusz and Surrente, Alessandro and Baranowski, Micha{\l} and Maude, Duncan K. and Malic, Ermin and Plochocka, Paulina},
  title = {Quadrupolar excitons in {MoSe$_2$} bilayers},
  journal = {Nat. Commun.},
  volume = {16},
  pages = {1382},
  year = {2025},
  doi = {10.1038/s41467-025-56586-3}
}

@article{Li2026DualGateWSe2,
  author = {Li, Zhe and Thor, Prokhor and Kourmoulakis, George and Ivanova, Tatyana V. and Taniguchi, Takashi and Watanabe, Kenji and Yu, Hongyi and Brotons-Gisbert, Mauro and Gerardot, Brian D.},
  title = {Highly tunable band structure in ferroelectric {R}-stacked bilayer {WSe$_2$}},
  journal = {Nat. Commun.},
  volume = {17},
  pages = {2457},
  year = {2026},
  doi = {10.1038/s41467-026-68854-x}
}

@article{togo2023first,
  author = {Togo, Atsushi},
  title = {First-principles Phonon Calculations with Phonopy and Phono3py},
  journal = {J. Phys. Soc. Jpn.},
  volume = {92},
  pages = {012001},
  year = {2023},
  doi = {10.7566/JPSJ.92.012001}
}



\clearpage
\onecolumngrid

\setcounter{page}{1}
\renewcommand{\thepage}{S\arabic{page}}
\renewcommand{\theHequation}{S\arabic{equation}}
\renewcommand{\theHfigure}{S\arabic{figure}}
\renewcommand{\theHtable}{S\arabic{table}}
\renewcommand{\theHsection}{S\arabic{section}}
\setcounter{equation}{0}
\setcounter{figure}{0}
\setcounter{table}{0}
\setcounter{section}{0}
\renewcommand{\theequation}{S\arabic{equation}}
\renewcommand{\thefigure}{S\arabic{figure}}
\renewcommand{\thetable}{S\arabic{table}}

\begin{center}
\textbf{\large Supplemental Material: Chiral-phonon-induced anomalous Hall effect in bilayer MoSe$_2$}\\[0.4cm]
Omer Yaniv$^{1}$, Swati Chaudhary$^{2}$, Carolina Paiva$^{1}$, and Dominik~M.~Juraschek$^{1,3,4,5}$\\[0.15cm]
{\small\itshape
$^1$School of Physics and Astronomy, Tel Aviv University, Tel Aviv 6997801, Israel\\
$^2$The Institute for Solid State Physics, The University of Tokyo, Kashiwa, Chiba 277-8581, Japan\\
$^3$Department of Applied Physics and Science Education, Eindhoven University of Technology, 5612 AP Eindhoven, Netherlands\\
$^4$Department of Physics, University of Fribourg, CH-1700 Fribourg, Switzerland\\
$^5$Paul Scherrer Institut, CH-5232 Villigen PSI, Switzerland
}
\end{center}

\section{First-principles extraction of the effective model parameters}

The low-energy Hamiltonian introduced in the main text is
\begin{equation}
H_Q(\mathbf{k})
=
\varepsilon_v(\mathbf{k})\tau_0
+
g(Q_y\tau_x-Q_x\tau_y)
+
h(Q_xk_y-Q_yk_x)\tau_z ,
\label{eq:supp_HQ}
\end{equation}
where $\mathbf{k}$ is measured relative to the $K$ valley. Here we
describe how the coupling constants $g$ and $h$ are extracted from
frozen-phonon first-principles calculations.

\subsection{Computational details}

We model centrosymmetric 2H-stacked bilayer MoSe$_2$ using a six-atom primitive slab cell containing two Mo and four Se atoms. The in-plane lattice constant is $a=3.288\,\text{\AA}$, and the out-of-plane cell length is $25\,\text{\AA}$. We perform first-principles calculations using density functional theory as implemented in VASP~\cite{kresse:1996,kresse2:1996}, including spin-orbit coupling, with the PBEsol exchange-correlation functional~\cite{perdew:2008}. Phonon frequencies and eigenvectors are obtained using Phonopy~\cite{togo2023first}. We use a plane-wave energy cutoff of $500\,\mathrm{eV}$ and an electronic convergence threshold of $10^{-8}\,\mathrm{eV}$, allowing up to $200$ electronic self-consistency iterations. The supercell calculations employ a $3\times3\times1$ supercell sampled with a $\Gamma$-centered $2\times2\times1$ $k$-point mesh.

\subsection{Extraction of the phonon-induced splitting $g$}

At the valley center, the momentum-dependent term in
Eq.~(\ref{eq:supp_HQ}) vanishes. For a frozen displacement along either
component of the $E_u$ phonon, the Hamiltonian therefore reduces to
\begin{equation}
H_Q(0)
=
\varepsilon_v(0)\tau_0
+
g(Q_y\tau_x-Q_x\tau_y).
\end{equation}
Its eigenvalue splitting is
\begin{equation}
S(K,Q)
\equiv
E_{48}(K,Q)-E_{47}(K,Q)
=
2g|Q|.
\label{eq:supp_g_split}
\end{equation}
We determine $g$ by performing frozen-phonon calculations for positive
and negative displacements along the two orthogonal components of the
$E_u$ mode and fitting the resulting splitting to
Eq.~(\ref{eq:supp_g_split}).

Figures~S1(a) and S1(b) show the small-amplitude behavior for
displacements along $Q_x$ and $Q_y$, respectively. Both components show
a linear splitting that is symmetric under reversal of the displacement.
The fits give
\begin{equation}
g_x=0.2939~
\frac{\mathrm{meV}}{\text{\AA}\sqrt{\mathrm{amu}}},
\qquad
g_y=0.2947~
\frac{\mathrm{meV}}{\text{\AA}\sqrt{\mathrm{amu}}},
\end{equation}
demonstrating the near equivalence of the two components of the
degenerate mode. We therefore use their common value
\begin{equation}
g\simeq0.294~
\frac{\mathrm{meV}}{\text{\AA}\sqrt{\mathrm{amu}}}
\end{equation}
in the effective model.

Figures~S1(c) and S1(d) extend the frozen-phonon calculations to larger
amplitudes. The $Q_x$ displacement remains close to the small-amplitude
linear behavior over the displayed range, whereas the $Q_y$ displacement
develops a pronounced nonlinear contribution beyond approximately
$Q\simeq0.8\,\text{\AA}\sqrt{\mathrm{amu}}$. This departure defines
the amplitude beyond which the leading order Hamiltonian used in the
main text is no longer quantitatively complete.

\begin{figure}[t]
    \centering
    \includegraphics[width=\columnwidth]{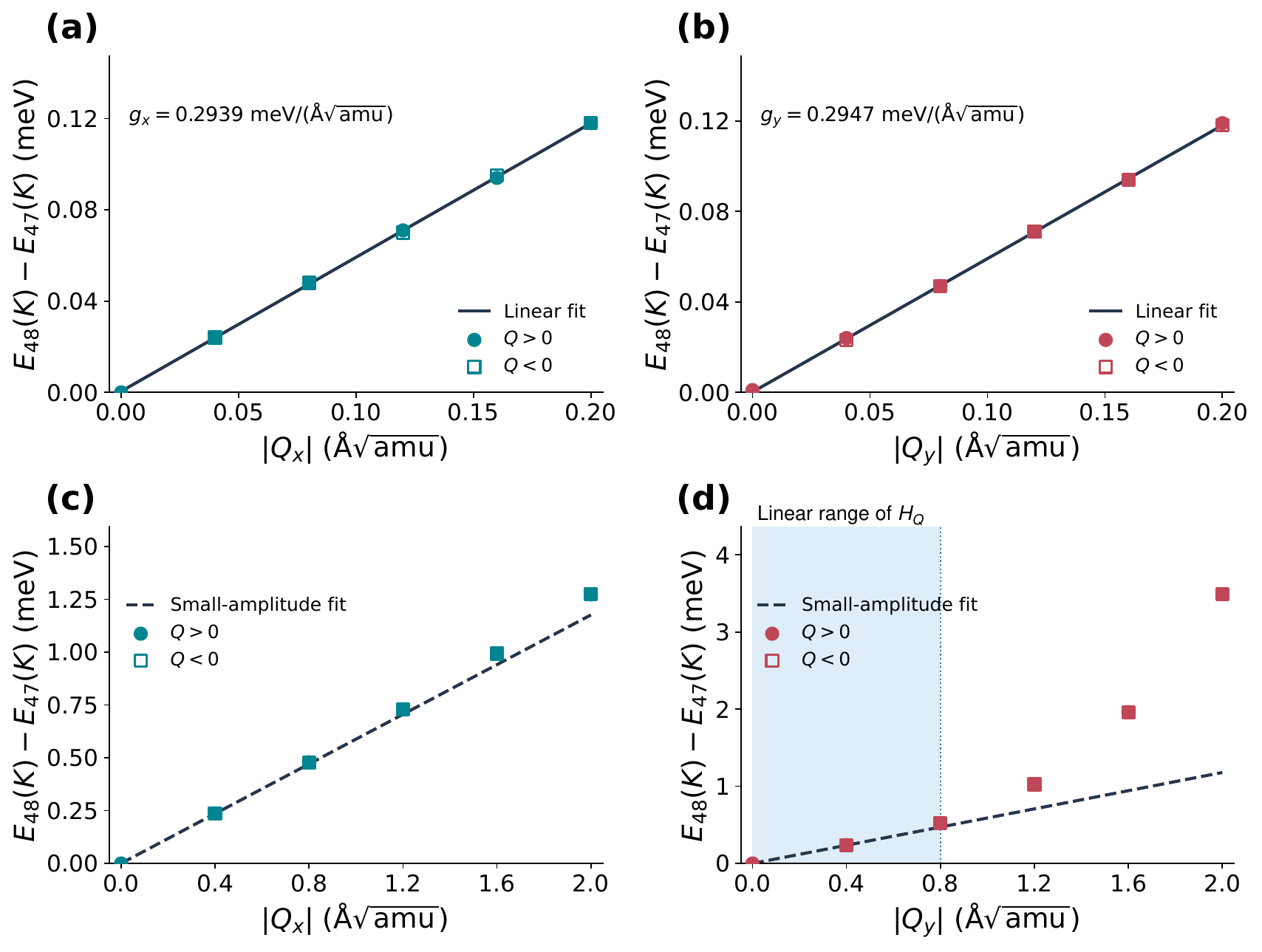}
    \caption{
    \textbf{Frozen-phonon extraction of the coupling $g$.}
    (a,b) Valence-band splitting
    $E_{48}(K)-E_{47}(K)$ as a function of the magnitude of the frozen
    phonon displacement along the two orthogonal components $Q_x$ and
    $Q_y$ of the $E_u$ mode. Filled and open symbols denote positive
    and negative displacements, respectively. Linear fits in the
    small-amplitude regime give
    $g_x=0.2939$ and
    $g_y=0.2947~\mathrm{meV}/(\text{\AA}\sqrt{\mathrm{amu}})$.
    (c,d) Splitting over an extended range of phonon amplitudes for
    $Q_x$ and $Q_y$, respectively. The dashed lines show the
    extrapolation of the small-amplitude linear fit. While the $Q_x$
    dependence remains close to linear, a pronounced nonlinear
    contribution develops for $Q_y$ beyond
    $Q\simeq0.8\,\text{\AA}\sqrt{\mathrm{amu}}$.
    }
    \label{fig:S1}
\end{figure}

\subsection{Extraction of the momentum-dependent coupling $h$}

The coefficient $h$ is obtained from the momentum dependence of the
phonon-induced splitting away from the valley center. Consider first a
frozen displacement along $Q_x$, such that $Q_x=Q$ and $Q_y=0$.
Equation~(\ref{eq:supp_HQ}) becomes
\begin{equation}
H_Q(k_y)
=
\varepsilon_v(k_y)\tau_0
-gQ\tau_y
+hQk_y\tau_z .
\end{equation}
The splitting of the two valence states is therefore
\begin{equation}
S(k_y,Q)
=
2|Q|\sqrt{g^2+h^2k_y^2}.
\label{eq:supp_split_h}
\end{equation}
At the valley center,
\begin{equation}
S(K,Q)=2g|Q|,
\end{equation}
so that Eq.~(\ref{eq:supp_split_h}) can be rearranged as
\begin{equation}
\frac{
\sqrt{S(k_y,Q)^2-S(K,Q)^2}
}
{2|Q|}
=
h|k_y|.
\label{eq:supp_h_fit}
\end{equation}
An analogous expression is obtained for a displacement along $Q_y$ and
a momentum cut along $k_x$.

Figure~S2 shows the quantity on the left-hand side of
Eq.~(\ref{eq:supp_h_fit}) obtained from the frozen-phonon band
structures. Its approximately linear momentum dependence provides a
direct extraction of $h$. The fits obtained from the different
displacement and momentum directions are mutually consistent, with
values close to
\begin{equation}
h\simeq 58~
\frac{\mathrm{meV}}{\sqrt{\mathrm{amu}}}.
\end{equation}
The spread between the different fits provides a measure of the
deviation from the minimal isotropic low-energy model.

At larger phonon amplitudes the momentum dependence also develops
higher-order corrections. In particular, the deviations visible for the
$Q_y$ displacement become substantially stronger than those for $Q_x$,
consistent with the nonlinear amplitude dependence found in
Fig.~S1. These results motivate restricting the quantitative
leading-order Floquet analysis of the main text to
$Q_0\lesssim0.8\,\text{\AA}\sqrt{\mathrm{amu}}$.

\begin{figure}[t]
    \centering
    \includegraphics[width=\columnwidth]{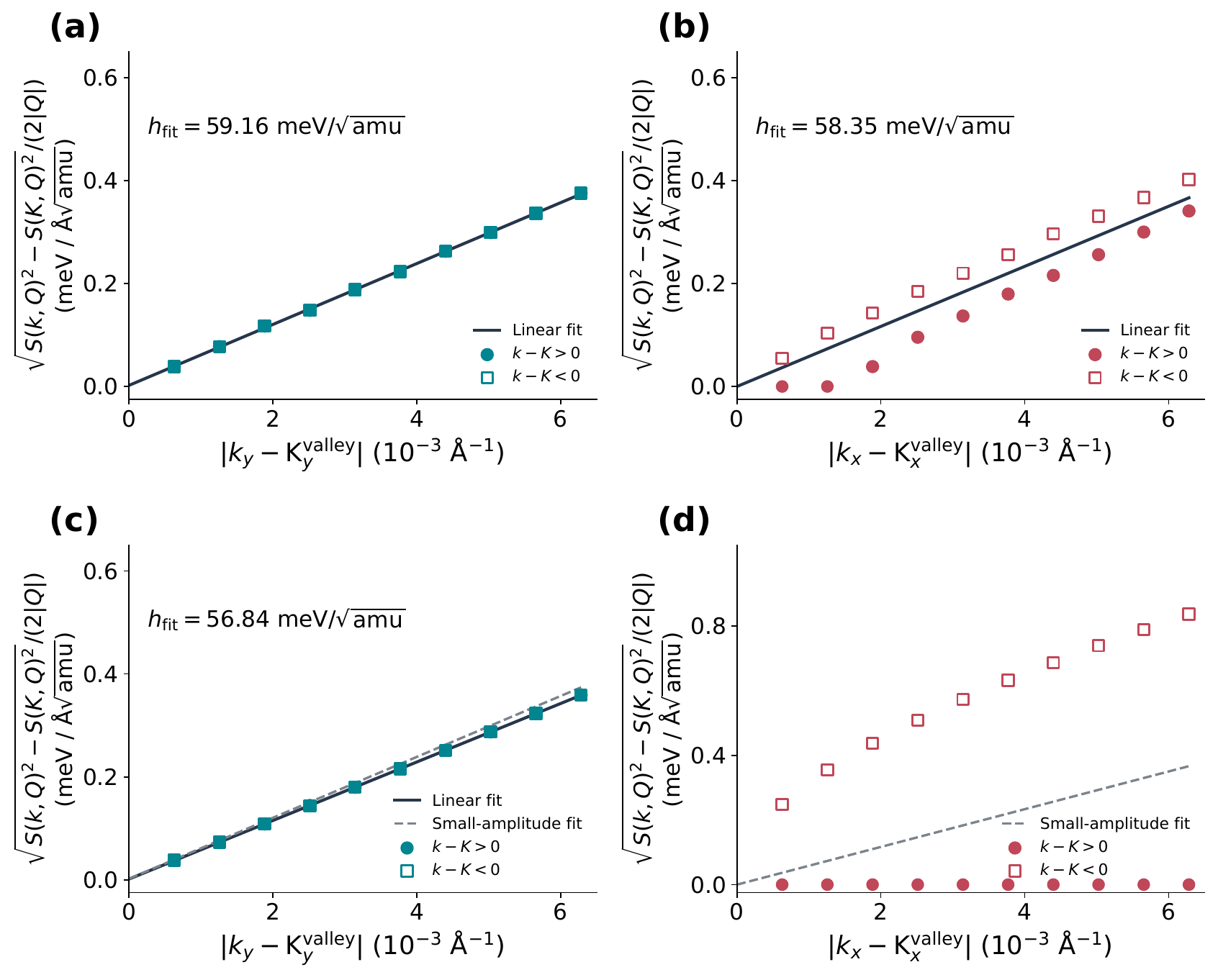}
    \caption{
    \textbf{Frozen-phonon extraction of the momentum-dependent coupling
    $h$.}
    The quantity
    $\sqrt{S(k,Q)^2-S(K,Q)^2}/(2|Q|)$ is plotted as a function of the
    momentum measured from the valley center. Within the leading-order
    Hamiltonian this quantity is equal to $h|k|$.
    (a,b) Momentum dependence for the two orthogonal combinations of
    phonon displacement and momentum direction in the small-amplitude
    regime. Filled and open symbols denote momenta on opposite sides of
    the $K$ valley. Linear fits give slopes of
    $59.16$ and $58.35~\mathrm{meV}/\sqrt{\mathrm{amu}}$, respectively.
    (c,d) Corresponding behavior at larger phonon amplitudes. Panel (c)
    remains close to the small-amplitude prediction, whereas panel (d)
    displays pronounced nonlinear and directional corrections.
    }
    \label{fig:S2}
\end{figure}

\section{Derivation of the effective Floquet Hamiltonian}

For circular excitation of the doubly degenerate $E_u$ phonon, we take
\begin{equation}
Q_x(t)=Q_0\cos\Omega t,
\qquad
Q_y(t)=Q_0\sin\Omega t.
\label{eq:supp_Qcirc}
\end{equation}
Substituting Eq.~(\ref{eq:supp_Qcirc}) into
Eq.~(\ref{eq:supp_HQ}) gives
\begin{align}
H(t)
={}&
\varepsilon_v(\mathbf{k})\tau_0
+
gQ_0
\left[
\sin(\Omega t)\tau_x
-
\cos(\Omega t)\tau_y
\right]
\nonumber\\
&+
hQ_0
\left[
k_y\cos(\Omega t)
-
k_x\sin(\Omega t)
\right]\tau_z .
\label{eq:supp_Htime}
\end{align}
We use the Fourier convention
\begin{equation}
H(t)
=
H_0+
H_{+1}e^{-i\Omega t}
+
H_{-1}e^{i\Omega t}.
\label{eq:supp_fourier}
\end{equation}
The static component is
\begin{equation}
H_0
=
\varepsilon_v(\mathbf{k})\tau_0 ,
\end{equation}
while the nonzero harmonics are
\begin{align}
H_{+1}
=
\frac{Q_0}{2}
\Big[
&i g\tau_x-g\tau_y
\nonumber\\
&+
h(k_y-i k_x)\tau_z
\Big],
\label{eq:supp_Hplus}
\end{align}
and
\begin{align}
H_{-1}
=
\frac{Q_0}{2}
\Big[
&-i g\tau_x-g\tau_y
\nonumber\\
&+
h(k_y+i k_x)\tau_z
\Big].
\label{eq:supp_Hminus}
\end{align}

In the regime
\begin{equation}
gQ_0,hQ_0k\ll\hbar\Omega,
\end{equation}
the leading correction to the time-averaged Hamiltonian is obtained from
the first-order van Vleck expansion,
\begin{equation}
H_F
=
H_0+
\frac{[H_{-1},H_{+1}]}{\hbar\Omega}.
\label{eq:supp_vV}
\end{equation}
Using
\begin{equation}
[\tau_i,\tau_j]
=
2i\epsilon_{ijk}\tau_k,
\end{equation}
the commutator becomes
\begin{equation}
[H_{-1},H_{+1}]
=
-Q_0^2
\left[
gh(k_x\tau_x+k_y\tau_y)
+
g^2\tau_z
\right].
\label{eq:supp_commutator}
\end{equation}
Therefore
\begin{equation}
H_F(\mathbf{k})
=
\varepsilon_v(\mathbf{k})\tau_0
-
\frac{Q_0^2}{\hbar\Omega}
\left[
gh(k_x\tau_x+k_y\tau_y)
+
g^2\tau_z
\right].
\label{eq:supp_HF}
\end{equation}
Dropping the constant $\varepsilon_v(0)$ and using
$\varepsilon_v(\mathbf{k})=\varepsilon_v(0)-ck^2$, we obtain
\begin{equation}
H_F(\mathbf{k})
=
-ck^2\tau_0
-
\frac{Q_0^2}{\hbar\Omega}
\left[
v(k_x\tau_x+k_y\tau_y)
+
\mu\tau_z
\right],
\label{eq:supp_HF_final}
\end{equation}
where
\begin{equation}
v=gh,
\qquad
\mu=g^2.
\end{equation}

Writing
\begin{equation}
H_F=d_0(\mathbf{k})\tau_0+
\mathbf{d}(\mathbf{k})\cdot\boldsymbol{\tau},
\end{equation}
with
\begin{equation}
\mathbf{d}(\mathbf{k})
=
-\frac{Q_0^2}{\hbar\Omega}
(vk_x,vk_y,\mu),
\label{eq:supp_dvector}
\end{equation}
the quasienergies are
\begin{equation}
E_\pm(k)
=
-ck^2
\pm
\frac{Q_0^2}{\hbar\Omega}
\sqrt{v^2k^2+\mu^2}.
\label{eq:supp_Epm}
\end{equation}
At the valley center this gives the Floquet gap
\begin{equation}
\Delta_F
=
E_+(0)-E_-(0)
=
\frac{2g^2Q_0^2}{\hbar\Omega}.
\label{eq:supp_gap}
\end{equation}

Reversing the phonon chirality corresponds to replacing
$Q_y(t)=Q_0\sin\Omega t$ by
$Q_y(t)=-Q_0\sin\Omega t$. This exchanges the two Fourier harmonics and
therefore reverses the sign of their commutator. The first-order Floquet
correction, including the mass term, consequently changes sign under
reversal of the circular phonon polarization.
\subsection{Berry curvature from the Floquet pseudospin texture}
For a generic two-band Hamiltonian of this form, the Berry curvature of
the two eigenstates is
\begin{equation}
\Omega_{\pm,z}(\mathbf{k})
=
\mp
\frac{1}{2}
\hat{\mathbf d}
\cdot
\left(
\partial_{k_x}\hat{\mathbf d}
\times
\partial_{k_y}\hat{\mathbf d}
\right),
\label{eq:supp_berry_general}
\end{equation}
where
\begin{equation}
\hat{\mathbf d}
=
\frac{\mathbf d}{|\mathbf d|}.
\end{equation}
Equivalently,
\begin{equation}
\Omega_{\pm,z}(\mathbf{k})
=
\mp
\frac{1}{2}
\frac{
\mathbf d\cdot
\left(
\partial_{k_x}\mathbf d
\times
\partial_{k_y}\mathbf d
\right)
}{
|\mathbf d|^3
}.
\label{eq:supp_berry_d}
\end{equation}
For the Floquet Hamiltonian of Eq.~(\ref{eq:supp_HF_final}),
\begin{equation}
\partial_{k_x}\mathbf d
=
-\frac{Q_0^2}{\hbar\Omega}
(v,0,0),
\end{equation}
and
\begin{equation}
\partial_{k_y}\mathbf d
=
-\frac{Q_0^2}{\hbar\Omega}
(0,v,0).
\end{equation}
Their cross product is
\begin{equation}
\partial_{k_x}\mathbf d
\times
\partial_{k_y}\mathbf d
=
\left(
\frac{Q_0^2}{\hbar\Omega}
\right)^2
(0,0,v^2).
\end{equation}
Using Eq.~(\ref{eq:supp_dvector}),
\begin{equation}
\mathbf d\cdot
\left(
\partial_{k_x}\mathbf d
\times
\partial_{k_y}\mathbf d
\right)
=
-
\left(
\frac{Q_0^2}{\hbar\Omega}
\right)^3
\mu v^2,
\end{equation}
while
\begin{equation}
|\mathbf d|^3
=
\left(
\frac{Q_0^2}{\hbar\Omega}
\right)^3
\left(
v^2k^2+\mu^2
\right)^{3/2}.
\end{equation}
The overall drive amplitude therefore cancels from the Berry curvature,
giving
\begin{equation}
\Omega_{\pm,z}(k)
=
\pm
\frac{1}{2}
\frac{\mu v^2}
{\left(v^2k^2+\mu^2\right)^{3/2}}.
\label{eq:supp_berry_full}
\end{equation}
Defining
\begin{equation}
\xi=\frac{\mu}{v}=\frac{g}{h},
\end{equation}
we obtain
\begin{equation}
\Omega_{\pm,z}(k)
=
\pm
\frac{1}{2}
\frac{\xi}
{\left(k^2+\xi^2\right)^{3/2}}.
\label{eq:supp_berry_xi}
\end{equation}
For the lower Floquet band,
\begin{equation}
\Omega_{-,z}(k)
=
-
\frac{1}{2}
\frac{\xi}
{\left(k^2+\xi^2\right)^{3/2}},
\end{equation}
for the chirality convention used above. Reversing the phonon chirality
reverses the sign of the Floquet correction and hence the sign of the
Berry curvature.

\section{Anomalous Hall conductivity}
\label{sec:supp_hall_derivation}

We evaluate the Hall conductivity at zero temperature, assuming Fermi occupations of the effective Floquet bands with chemical potential $E_F=E_-(0)$. Defining $A=Q_0^2/(\hbar\Omega)$, the quasienergies are
\begin{equation}
E_\pm(k)=-ck^2\pm A\sqrt{v^2k^2+\mu^2},
\end{equation}
where $v=gh$, $\mu=g^2$, and $\xi=\mu/v=g/h$. We use positive $c$, $g$, $h$, and $\Omega$, and a nonzero drive amplitude.

The lower band satisfies $E_-(k)\le E_-(0)$ and is therefore occupied throughout the momentum range considered. The upper band is unoccupied near the valley center and becomes occupied beyond the boundary $k_F$ defined by
\begin{equation}
E_+(k_F)=E_-(0)=-A\mu.
\end{equation}
Consequently,
\begin{align}
ck_F^2
&=A\left[\sqrt{v^2k_F^2+\mu^2}+\mu\right]\nonumber\\
&=Av\left[\sqrt{k_F^2+\xi^2}+\xi\right].
\label{eq:supp_occupation_boundary}
\end{align}
Introducing $s=\sqrt{k_F^2+\xi^2}$, so that $k_F^2=s^2-\xi^2$, gives
\begin{equation}
c(s-\xi)(s+\xi)=Av(s+\xi).
\end{equation}
Since $s+\xi>0$, this reduces to
\begin{equation}
s=\xi+\frac{Av}{c}
=\xi\left(1+\frac{Av}{c\xi}\right).
\end{equation}
Using $v/\xi=(gh)/(g/h)=h^2$, we obtain
\begin{equation}
\sqrt{k_F^2+\xi^2}=\xi\left[1+\beta(Q_0)\right],
\qquad
\beta(Q_0)=\frac{h^2Q_0^2}{c\hbar\Omega}.
\label{eq:supp_kF_beta}
\end{equation}
Squaring this expression yields
\begin{equation}
k_F^2=\xi^2\left[(1+\beta)^2-1\right]
=\xi^2\beta(2+\beta).
\end{equation}

The two Floquet bands have opposite Berry curvatures, $\Omega_{+,z}=-\Omega_{-,z}$. Their contributions therefore cancel wherever both bands are occupied, leaving only the disk $k<k_F$ around each valley. Writing the lower-band curvature as
\begin{equation}
\Omega_{-,z}(k)=\chi\frac{\xi}{2(k^2+\xi^2)^{3/2}},
\end{equation}
where $\chi=\pm1$ specifies its sign for the chosen phonon handedness ($\chi=-1$ for the convention in the preceding derivation), the contribution from one valley is
\begin{align}
\sigma_{xy}^{K}
&=-\frac{e^2}{\hbar}\int_0^{k_F}\frac{k\,dk}{2\pi}\,\Omega_{-,z}(k)\nonumber\\
&=-\chi\frac{e^2}{2\pi\hbar}
\left[1-\frac{\xi}{\sqrt{k_F^2+\xi^2}}\right],
\end{align}
 Here we used
\begin{equation}
\int_0^{k_F}\frac{k\xi\,dk}{(k^2+\xi^2)^{3/2}}
=1-\frac{\xi}{\sqrt{k_F^2+\xi^2}}.
\end{equation}

The two valleys contribute with the same sign. Including both and substituting Eq.~(\ref{eq:supp_kF_beta}) therefore gives
\begin{equation}
\sigma_{xy}(Q_0)=-\chi\frac{e^2}{h_{\rm P}}
\frac{\beta(Q_0)}{1+\beta(Q_0)},
\end{equation}
and hence
\begin{equation}
\boxed{\left|\sigma_{xy}(Q_0)\right|
=\frac{e^2}{2\pi\hbar}\frac{\beta(Q_0)}{1+\beta(Q_0)}}.
\label{eq:supp_hall_vs_amplitude}
\end{equation}

Using the Floquet gap
$\Delta_F=2g^2Q_0^2/(\hbar\Omega)$, we have
$\beta(Q_0)=h^2\Delta_F/(2cg^2)$.
Equation~(\ref{eq:supp_hall_vs_amplitude}) can therefore
be expressed directly in terms of the Floquet gap as
\begin{equation}
\boxed{
\left|\sigma_{xy}\right|
=
\frac{e^2}{2\pi\hbar}
\frac{h^2\Delta_F}{2cg^2+h^2\Delta_F}
}.
\label{eq:supp_hall_vs_gap}
\end{equation}

\end{document}